%% file: Manuscript.tex
\documentclass[preprint, 10 pt]{elsarticle}
\usepackage{natbib}
\biboptions{authoryear}

\usepackage{graphicx}
\usepackage{hyperref}
\usepackage[utf8]{inputenc}
\usepackage{multirow}
\usepackage{array,multirow}
\usepackage{hyperref}
\usepackage{caption}
\usepackage{subcaption}
\usepackage[legalpaper, margin=1in]{geometry}
\usepackage[table]{xcolor}

\usepackage{amssymb}
\usepackage{amsmath}
\usepackage{xcolor}

\journal{arXiv}
\begin{document}

\begin{frontmatter}


\title{Multiclass Semantic Segmentation to Identify Anatomical Sub-Regions of Brain and Measure Neuronal Health in Parkinson’s Disease}

 \author[address1,address2]{Hosein Barzekar}
\ead{barzekar.h@gmail.com}

\author[address3]{Hai Ngu}
\ead{hain@gene.com}

\author[address1]{Han Hui Lin}
\ead{hanhlin@gene.com}

\author[address2]{Mohsen Hejrati}
\ead{hejratis@gene.com}

\author[address1]{Steven Ray Valdespino}
\ead{srvaldespino94@gmail.com}

\author[address1]{Sarah Chu}
\ead{chus18@gene.com}

\author[address1]{Baris Bingol}
\ead{barisb@gene.com}

\author[address2]{Somaye Hashemifar\corref{corresponding author}}
\cortext[corresponding author]{Corresponding author}
\ead{hashems4@gene.com}

\author[address1]{Soumitra Ghosh \corref{corresponding author}}
\cortext[corresponding author]{Corresponding author}
\ead{ghoshs29@gene.com}

\address[address1]{Department of Neuroscience, Genentech Inc., South San Francisco, CA, USA}
\address[address2]{Department of Artificial Intelligence, Genentech Inc., South San Francisco, CA, USA}

\address[address3]{Department of Pathology, Genentech Inc., South San Francisco, CA, USA}

\begin{abstract}
Automated segmentation of anatomical sub-regions with high precision has become a necessity to enable the quantification and characterization of cells/ tissues in histology images. Currently, a machine learning model to analyze sub-anatomical regions of the brain to analyze 2D histological images is not available. The scientists rely on manually segmenting anatomical sub-regions of the brain which is extremely time-consuming and prone to labeler-dependent bias. One of the major challenges in accomplishing such a task is the lack of high-quality annotated images that can be used to train a generic artificial intelligence model. In this study, we employed a UNet-based architecture, compared model performance with various combinations of encoders, image sizes, and sample selection techniques.  Additionally, to increase the sample set we resorted to data augmentation which provided data diversity and robust learning. In this study, we trained our best fit model on approximately one thousand annotated 2D brain images stained with Nissl/ Haematoxylin and Tyrosine Hydroxylase enzyme (TH, indicator of dopaminergic neuron viability). The dataset comprises of different animal studies enabling the model to be trained on different datasets. The model effectively is able to detect two sub-regions compacta (SNCD) and reticulata (SNr) in all the images. In spite of limited training data, our best model achieves a mean intersection over union (IOU) of 79\% and a mean dice coefficient of 87\%. In conclusion, the UNet-based model with EffiecientNet as an encoder outperforms all other encoders, resulting in a first of its kind robust model for multiclass segmentation of sub-brain regions in 2D images.

\end{abstract}


\end{frontmatter}

\input{01_Introduction.tex}

\input{03_Methodology}

\input{04_Results.tex}

\input{05_Discussion.tex}

\bibliographystyle{apalike}
\bibliography{References}

\end{document}

%% file: 01_Introduction.tex
\section{Introduction}

Deep learning-based image segmentation approaches for medical imaging are gaining interest due to their self-learning and generalization ability \cite{kayalibay2017cnn}. For instance, deep learning semantic segmentation models can facilitate robust and fast approaches to detect boundaries between different Regions Of Interest (ROI) in histopathology images, enabling more sensitive analysis of biological experiments in animals and humans\cite{Moshkov2020Test}. Analysis of medical images to interpret the disease-associated pathology typically requires the identification of specific regions by an expert pathologist, and followed by segmentation and annotation that is done manually. It is a tedious and time-consuming task which is also susceptible to inter- and intra-observer variability. Although image segmentation of natural images has attained a high level of performance in the field of machine learning, the strict segmentation requirements for histopathology images are yet to be achieved. Penttinen et al. \cite{penttinen2018implementation} shows the latest deep learning model to quantify dopaminergic neurons in the entire mouse brain. Although a step forward towards automated analysis of dopaminergic (DA) neurons health in Parkinson's Disease (PD), it lacks the ability to compartmentalize the sub-anatomical regions of the Substantia Nigra (SN). To measure the ability of a potential drug in abrogating PD progression, it is very essential to assess the dopaminergic neuronal health in SNr followed by SNCD. The loss of dopaminergic neurons often starts in the SNr region where most of the axons for DA neurons reside. The degeneration from the axons extend all the way to the neuronal cell body and nucleus which is mainly located in the SNCD region \cite{ghosh2021alpha, fu2012cytoarchitectonic}. Thorough analysis of DA neuronal health in these two regions is pivotal to assess the potential of new drugs in mitigating the disease progression. Despite demand, little progress has been made because of the difficult design requirements and lack of large-scale high quality datasets.

Over the past few decades, numerous semantic segmentation models have emerged in response to ever-rising technical demands. These algorithms have become essential in many visual-based applications that enable a granular understanding of the images including remote sensing \cite{hossain2019segmentation}, automated driving\cite{kaymak2019brief}, and medical image analysis\cite{zhuang2019application, roth2018deep}. The well-known algorithms for semantic segmentation include Fully Convolutional Network (FCN), SegNet, PSPNet, UNet, and DeepLab 
\cite{long2015fully, badrinarayanan2017segnet, zhao2017pyramid, ronneberger2015u, chen2017deeplab}. 
FCN uses the fully convolutional neural network for the end-to-end training of segmentation. It modifies the structure of VGG-16 and other networks by replacing every fully connected layers with convolutional layers \cite{long2015fully}. The most important characteristics of FCN are upsampling and skip connections which enables the model to expand the convolution image to the size of the original image without restoring the original value. Both SegNet and PSPNet are based on FCN. SegNet for the first time utilizes symmetric network structure \cite{badrinarayanan2017segnet} and PSPNet introduces the pyramid pooling module to fuse hierarchical scale features \cite{zhao2017pyramid}. To improve on previous results achieved by FCN, a new architecture called UNet based on an encoder-decoder structure was proposed. \cite{ronneberger2015u}. Each encoder module includes a sequence of convolutional layers followed by a pooling layer and each decoder module includes a concatenation of the output feature maps of the correspondent encoder module, sequence of convolutions and an upsampling layer \cite{ronneberger2015u}. UNet++ is a nested UNet architecture to improve the efficacy for medical image segmentation. UNet++ borrows the dense connection of DenseNet and re-designs the skip connection structure of UNet to reduce the semantic gap between the feature maps of the encoder and decoder sub-networks \cite{zhou2018unet++}. The VNet has a similar structure as UNet and is developed for analyzing 3D images in CT and MRI images \cite{milletari2016v}. DeepLabv1 combines atrous convolution with CNN to explicitly control the resolution at which feature responses are computed within Deep Convolutional Neural Networks \cite{chen2017deeplab}. To optimize performance, DeepLabv2 introduces atrous spatial pyramid pooling (ASPP), which utilized atrous convolution to get multi-scale information. DeepLabv3 improved the ASPP model with utilizing one 1$\times$1 convolution and three 3$\times$3 convolution \cite{chen2018encoder}.

In this study, we establish a deep learning-based framework for segmentation and quantification of two sub-regions of the SN in mouse brain tissues.. The SN is the most suseptible region to dopmainergic neuronal loss in PD. PD is the second most common neurodegenerative disease in the world and currently affects 100 million people in US. The two specific sub-regions of SN that are highly sensitive to genetic and sporadic stressors are Substantia Reticular part (SNr) and Substantia Nigra Compacta, dorsal tier (SNCD). Accurate and precise segmentation of sub-regions of SN is key to understanding the processes underlying PD progression \cite{menke2010connectivity}. The SNr and SNCD is encompassed with dopaminergic (DA) neurons, a unique type of neurons that expresses an enzyme Tyrosine Hydroxylase (TH). Loss of these neurons as indicated by reduced TH signal 
 is associated with motor neuron deficits and symptoms associated with PD. Every pre-clinical study to understand PD pathology or efficacy of a new potential drug requires careful analysis of TH signal in SNr and SNCD. It is challenging to analyze the TH signal in SNr and SNCD due to their small size and absence of visual cues to human eyes. It is mainly detected by TH staining or other staining that detects dopaminergic neurons in the SN region. There are a few studies available for the automatic segmentation of SN on MRI images of human brains, which are mostly based on brain atlas, template, and prior shape model. A model of such sort does-not take into account the age and sex of the animal which can heavily influence the size of the brain which in turn affects the size and location of these sub-regions in the brain \cite{haegelen2013automated, visser2016automated, guo2018seed, garzon2018automated, basukala2021automated}. To the best of our knowledge, this is the first study of developing a deep convolution neural network on 2D pathology images to segment sub-regions of interest for a granular understanding of the disease progression.  

Given the complexity of digital pathology images, our model employs an encoder-decoder paradigm by applying a transfer learning-based model to serve as a feature extractor in the encoder phase. Furthermore, a comprehensive analysis of the performance of several models as feature extractors was conducted. Finally, experimenting with different augmentations and image sizes, the output data clearly shows that the model's performance is on par with a human expert while being accomplished in a remarkably short amount of time using a relatively small amount of data in an unbiased manner.

In summary, we make the following contributions:

\begin{itemize}
    \item This approach is the first of its kind to automatically segment regions of interest (ROIs) from histopathological images of PD.
    \item Training of the model is conducted across a wide range of encoders, augmenting techniques, and image sizes. 
    \item Extensive experiments showing that the model can detect ROIs using a relatively small amount of labeled data. The final model is further tested on a blind test dataset (a separate mouse study that the model has not been trained on and not part of the train/validation/test set used for model development).
    \item The post-processing step to measure the TH staining optical density also provides the health of dopaminergic neurons in the specific ROIs which is critical to assess the pathology in PD.
\end{itemize}

%% file: 03_Methodology.tex
\section{Methodology}
\subsection{Dataset}

The dataset used in this study was obtained from different pathological studies (multiple internal datasets) where mouse brains were sectioned at 35 micron thickness and stained with TH and either Haematoxylin or Nissl as a background tissue stain. The sections were then imaged using a whole slide scanner microscope, Nanozoomer system (Hamamatsu Corp, San Jose, CA) at 20x resolution (0.46 microns/pixel). Whole coronal brain section images containing the SN were exported from the digital scans at 5x resolution (1.84 microns/pixels) and were used to train the model. This procedure helped us to obtain approximately one thousand 2D mouse brain images that have both SNr and SNCD. The ground truth (GT) for this study was drawn by biologists who specialize in mouse brain anatomy. 
Specifically, for the blind test dataset (a separate mouse study in which the model has not been directly trained on) as shown in \autoref{fig:TH_OD}, the animals in (c) and (d) were injected with a virus (viral over-expression of a protein that causes intentional dopaminergic neuronal loss in one side of the brain under experimental conditions) in the SN to induce loss of dopaminergic neurons, as indicated by the decrease in TH signal in SNr and SNCD.

\subsection{Preprocessing}
\begin{figure}
    \centering
    \includegraphics[width=.8\linewidth, height=8cm]{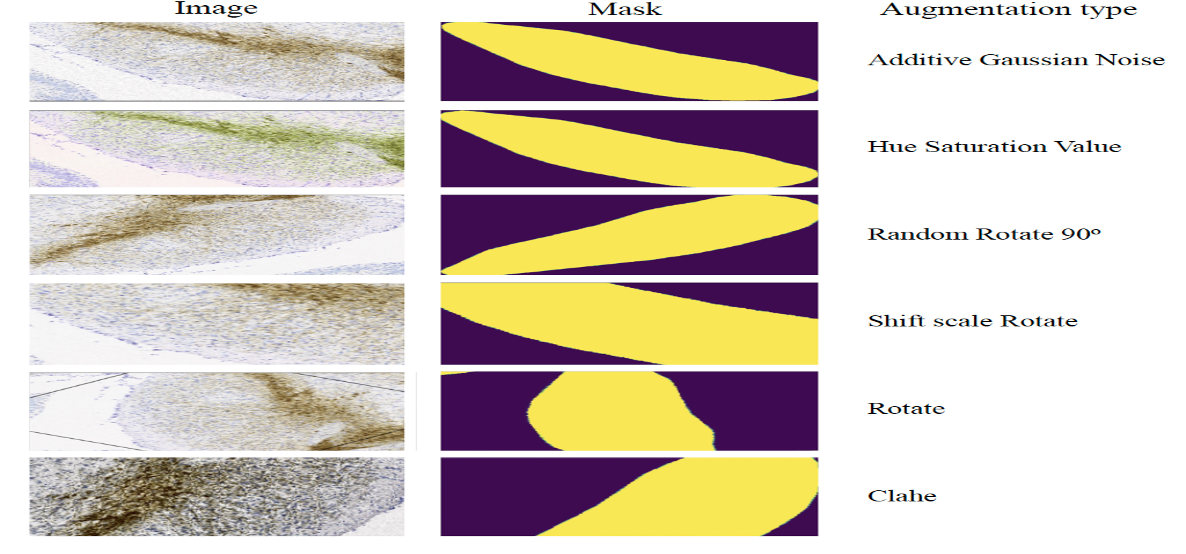}
    \caption{Various image augmentation techniques used in by our model}
    \label{fig:aug}
\end{figure}

Original images come in different sizes. First, each image is downsized to 1024 by 1024 pixels.
Deep learning models are notoriously susceptible to overfitting; hence, augmentation techniques are applied to the images to prevent or mitigate this occurrence as shown in \autoref{fig:aug}. The albumentation \cite{info11020125} library is utilized to reach our objective. The input images have undergone rotation, vertical flip, horizontal flip, random 90-degree rotation, transposition, elastic transformation, and the addition of gaussian noise.

\subsection{Methods}

Encoder-decoder architecture is used to deploy the model. This structure has been the source of many well-known model including the UNet \cite{ronneberger2015u}. The encoder is the feature extractor of the model and generates feature maps from the input images. For deep learning algorithms to work properly, training and future data should indeed coexist in the same space, and the more data it has access to, the better it performs. Because of the scarcity of data in many fields, including medicine, transfer learning is a useful tool for filling in the gaps. As a result, pre-trained networks can be used to acquire some of the fundamental parameters \cite{pan2009survey}. Hence, EfficientNet \cite{tan2019efficientnet} is used as to serve as the encoder for feature extraction. A new approach to scaling was introduced by EfficientNet, which leveraged a straightforward but exceedingly practical compound coefficient to equally scale depth, width, and resolution. EfficientNet  has different models available which also considers multiple parameters into account for each input. After experimenting with multiple architectures and their performance, EfficientNet-B5 is used as the encoder. Number of parameter and FLOPs for this model are 30M and 9.9B, respectively. UNet is used as the decoder where feature maps generated by the encoder served as the input to go through upsampling layers. Since the dataset involves multiclass segmentation, softmax \autoref{eqSoftmax} activation function is used in the final layer.

\begin{equation}
\label{eqSoftmax}
\begin{split}
\sigma(\Vec{\mathbf{x}})_i =\frac{e^{x_i}}{\sum_{j=1}^C e^{x_j}}
\end{split}
\end{equation}

where $\Vec{\mathbf{x}}$ is the input vector and C is the number of classes.

The loss function for employing the architecture is a Dice \autoref{eqlossFunction} function represented as follow: 

\begin{equation}
\label{eqlossFunction}
\begin{split}
Dice = \frac{2 *\sum_{i=1}^{N}y_{i}\hat{y_{i}}}
{\sum_{i=1}^{N}y_{i} + \sum_{i=1}^{N}\hat{y_{i}}}
\end{split}
\end{equation}

where  $y_i$ is the actual label on $\hat{y}_i$ is the predicted label and $N$ is the number of classes, which in our case is two, including SNr and SNCD.

\subsection{TH Intensity}
The TH signal was analyzed using MATLAB v9.12 (MathWorks, Natick, MA). TH positive stain pixel area was determined using color thresholds and a blue normalization algorithm \cite{Brey} at 20x resolution. The optical density \autoref{eqOpticalDensity} was calculated for the positive pixels according to Beer-Lambert law:

\begin{equation}
\label{eqOpticalDensity}
\begin{split}
OD = -log(\frac{I}{255})
\end{split}
\end{equation}

where $I$ is the pixel 8-bit grayscale intensity (in the range of 1 to 255; in the case of 0, $I$ is estimated to be 1). The OD of the positive pixels within the region of interest (manual ground truth region or prediction region generated by the model) was summed and normalized to the region area.

%% file: 04_Results.tex
\section{Results} 
\subsection{Setup}
Here we show the outcomes of the entire study in which approximately a thousand histopathological images encompassing SNr SNCD stained with TH has been split into training, validation, and a test set at percentages of 80\%, 10\%, and 10\%, respectively.

\subsection{Hyperparameter Tuning}
Model parameters are fine-tuned when the encoder is initialized using the pre-trained models. Specifically, Adam is used as the optimizer with the learning rate set to $1{e}^{-3}$ and  $\beta_1$, $\beta_2$  to 0.9 and 0.999, respectively.  As a learning rate decay scheduler, ReduceLROnPlateau is employed. The model is trained for 50 epochs using a mini-batch gradient descent with different batch sizes including 2, 4, and 8. To avoid overfitting, an early stopping mechanism is utilized on the validation set.

Summary of all the parameters and hyperparameters used for the model is shown in \autoref{hyperparameter}.

\begin{table}[htbp]
  \centering

  \caption{{summary of all the parameters and hyperparameters used in the model}} 
    \begin{tabular}{cccc}
    \hline
    Loss & Optimizers & Learning Rate & Image Sizes \\ \hline
    Dice  & Adam, SGD  & $10^{-3}$ - $10^{-6}$ & 512, 768, 1024 \\
    Jaccard  & Adam, SGD   & $10^{-3}$ - $10^{-6}$  &  512, 768, 1024 \\
    Categorical cross-entropy  & Adam, SGD & $10^{-3}$ - $10^{-6}$  & 512, 768, 1024 \\ \hline
    \end{tabular}%

  \label{hyperparameter}%
\end{table}%

\subsection{Model Selection}

Initially, we evaluate the performance of two popular architectures UNet and DeepLabv3+ with
both ResNet-50 model as its backbone. Furthur investigation on DeepLabv3+ was stopped due to lower performance compare to UNet (10\% lower than UNet model using validation set, data not shown). The UNet style model was selected for our experiments. Six different backbones including VGG-19, ResNet-50, ResNet-34, DenseNet-121, EfficientNet, and MobileNet are then employed \cite{Yakubovskiy:2019}. IOU score is the metric in which the model's performance is measured. In addition, all the models utilize Dice loss as the loss function. We consider training the target model using a transfer learning approach from the standard supervised pre-trained model on ImageNet, which is the de facto transfer learning pipeline in medical imaging.

\begin{figure}
    \centering
    \includegraphics[width=0.8\linewidth]{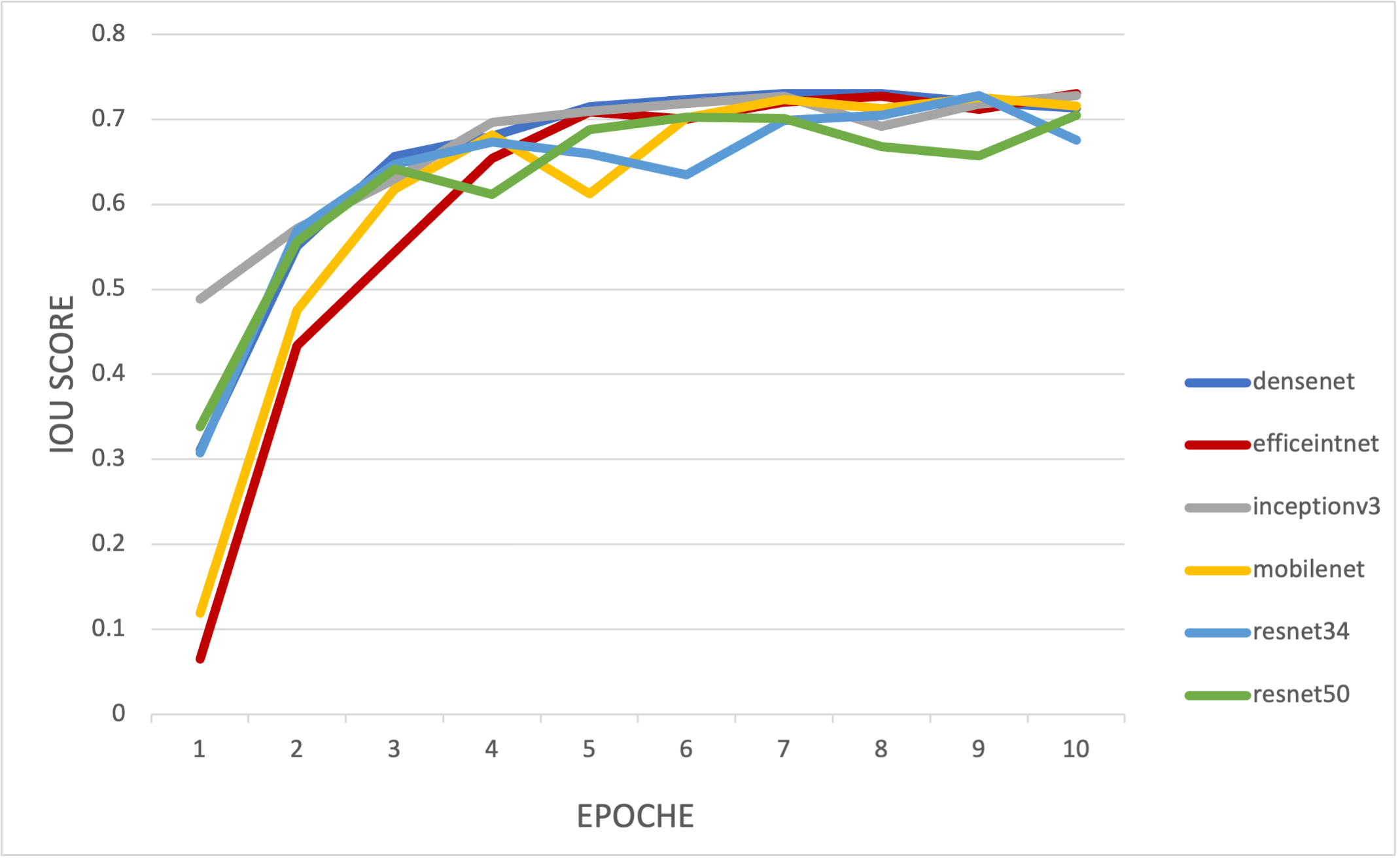}
    \caption{Model selection including six different networks as backbone}
    \label{fig:modelselection}
\end{figure}

\begin{figure}
    \centering
    \includegraphics[width=0.8\linewidth]{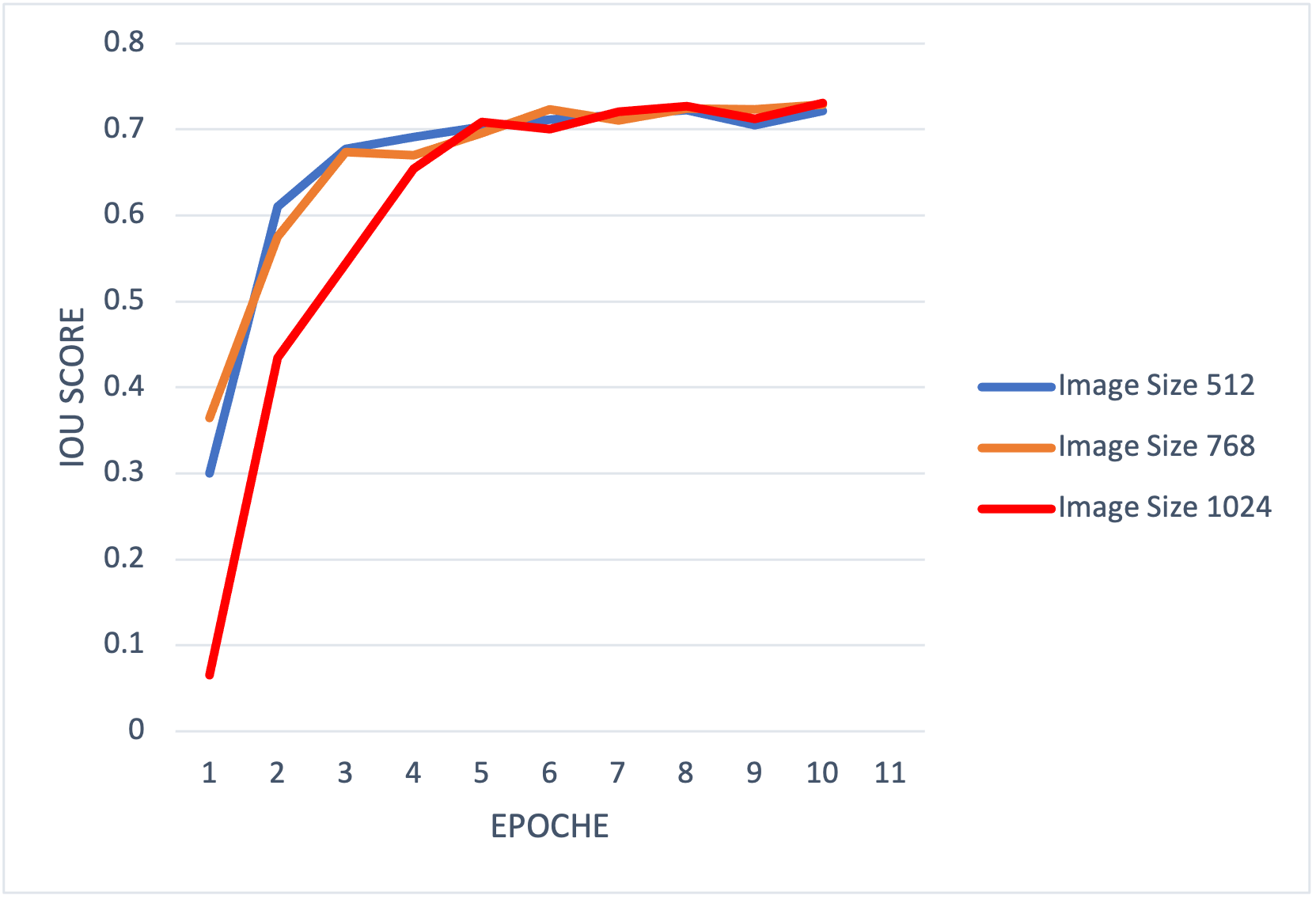}
    \caption{Investigating Image size on the model}
    \label{fig:imagesize}
\end{figure}

\autoref{fig:modelselection} shows the result of different backbones used as the encoder of the UNet model from which we draw the following conclusion: UNet with EfficientNet backbone and transfer learning from the ImageNet database provides better performance (0.73 IOU) compared with other network backbones. Furthermore, we investigate image size to choose the proper size for our experiments. \autoref{fig:imagesize} shows the result of the different image sizes having Efficientb5 as the encoder and UNet as decoder. An image size of 1024$\times$1024 resulted in better performance.

\subsection{Evaluation Metrics}

To further evaluate the performance of the model, the following metrics are also employed. \\
TP, FP, and FN in below equations, are true positive, false positive, and false negative, respectively. \\
\textbf{\textit{Precision}}: Precision, \autoref{eqPrecision}, indicates of all the pixels inside the predicted segmentation area what fraction of them are accurately segmented.
\begin{equation}
\label{eqPrecision}
\begin{split}
Precision =\frac{TP}{TP+FP}
\end{split}
\end{equation} \\
\textbf{\textit{Recall}}: Recall, \autoref{eqRecall}, measures the proportion of successfully separated pixels inside the ground truth area. 

\begin{equation}
\label{eqRecall}
\begin{split}
Recall =\frac{TP}{TP+FN}
\end{split}
\end{equation} \\

\textbf{\textit{Dice coefficient}}: The Dice or F1 score, \autoref{eqDiceCoefficient}, is frequently used to evaluate the performance of image segmentation models that measure the degree to which two semantic masks are identical. Therefore, it is the size of the overlap between the two segmentations divided by the combined size of the two segmentations.

\begin{equation}
\label{eqDiceCoefficient}
\begin{split}
\textit{Dice coefficient} =\frac{2 \times TP}{2  \times  TP + FP + FN}
\end{split}
\end{equation} \\

\subsection{Quantitative and Qualitative results}

 Based on our results,  the UNet-EfficientNet combination was used as the model with an image input size of 1024$\times$1024.

 \begin{table}[htbp]
  \centering

  \caption{{Performance of the model on the test set with \& without elastic transformation(ET)}} 
    \begin{tabular}{ccc}
    \hline
    Evaluation Metric & Mean value w/ ET & Mean value w/o ET \\ \hline
    IOU Score  & 79\% & 78\% \\
    Dice Coefficient & 87\% & 86\% \\
    Recall  & 88\% &  87\%\\
     Precision  & 86\% & 85\%\\ \hline
    \end{tabular}%

  \label{resultET}%
\end{table}%

\autoref{resultET} shows the result of the model on the test set. The model achieves a Dice coefficient of 0.87 on the multiclass segmentation task. Considering the complexity and variability of the staining and different morphologies in different mice, the model is considered to be performing well after qualitative review by expert biologists. Results in \autoref{resultET} indicate that including elastic transformation boosted the performance by 1\% on the metrics.

\begin{figure}
        \begin{subfigure}[b]{.2\textwidth}
                \includegraphics[width=.75\linewidth]{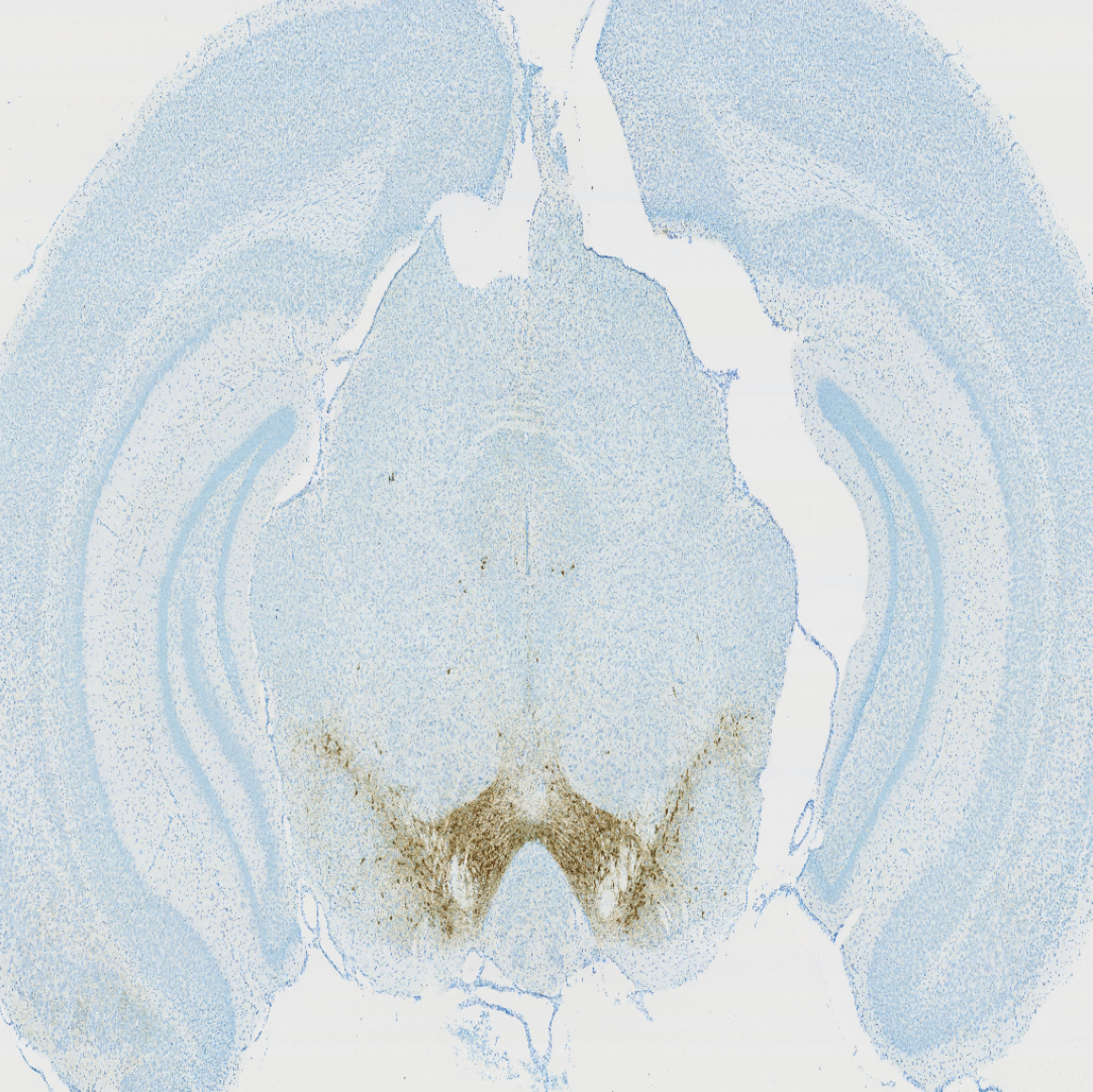}
                \caption{Image 1}
                \label{fig:image_test_1}
        \end{subfigure}%
        \begin{subfigure}[b]{.2\textwidth}
                \includegraphics[width=.75\linewidth]{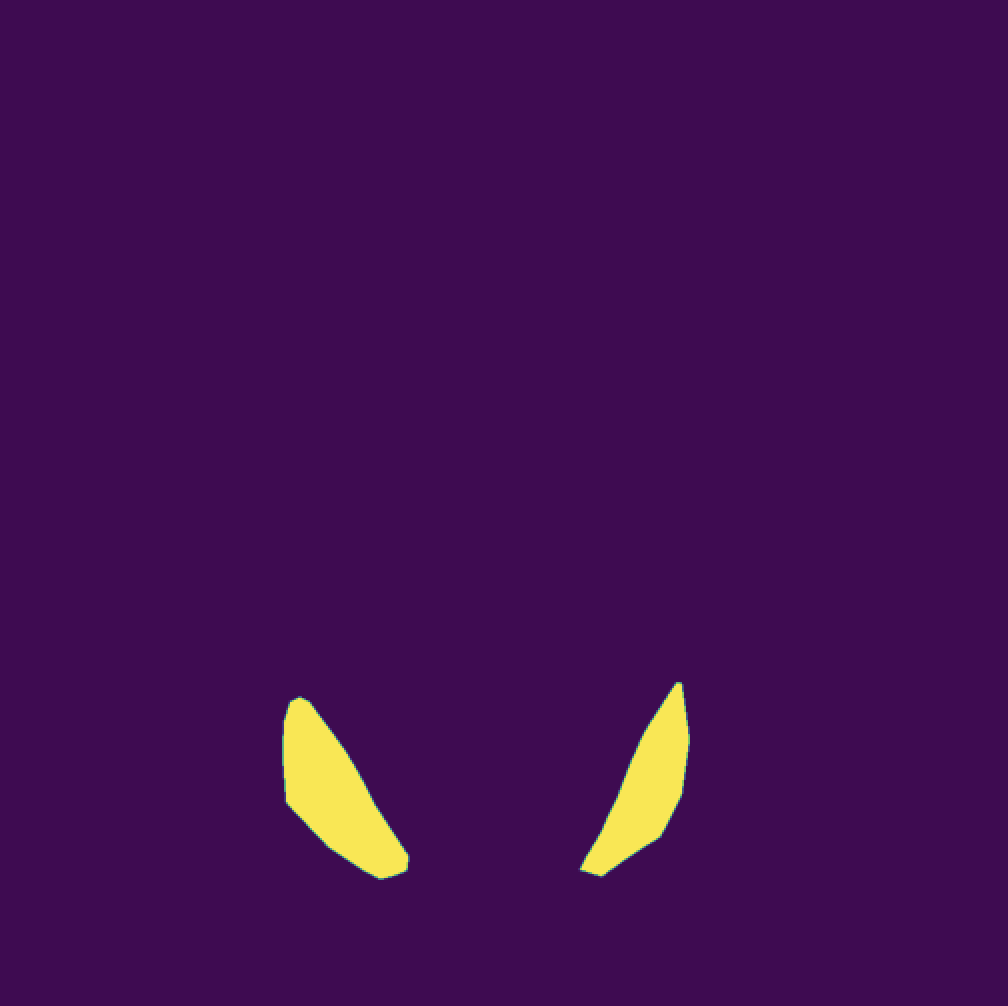}
                \caption{GT SNr}
                \label{fig: SNr_gt}
        \end{subfigure}%
        \begin{subfigure}[b]{.2\textwidth}
                \includegraphics[width=.75\linewidth]{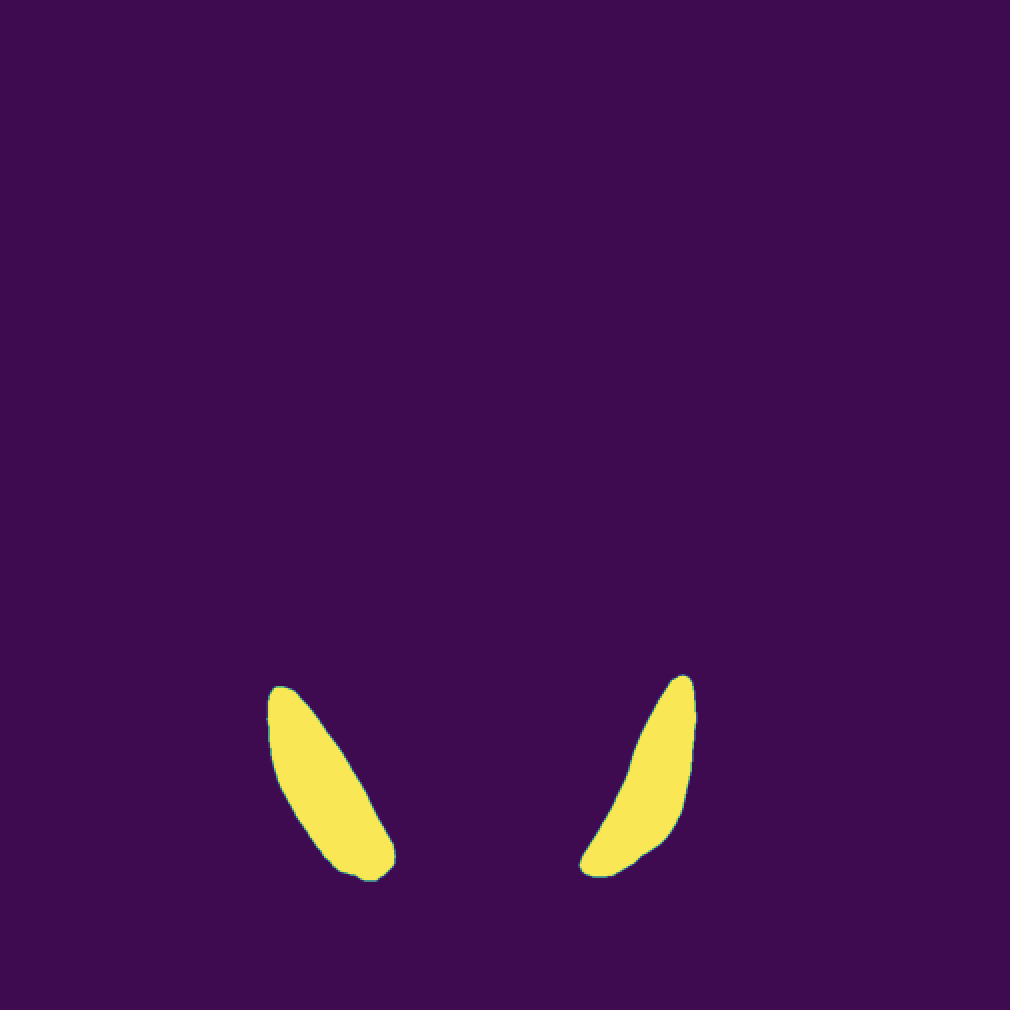}
                \caption{pred SNr}
                \label{fig:pred_SNr}
        \end{subfigure}%
        \begin{subfigure}[b]{0.2\textwidth}
                \includegraphics[width=.75\linewidth]{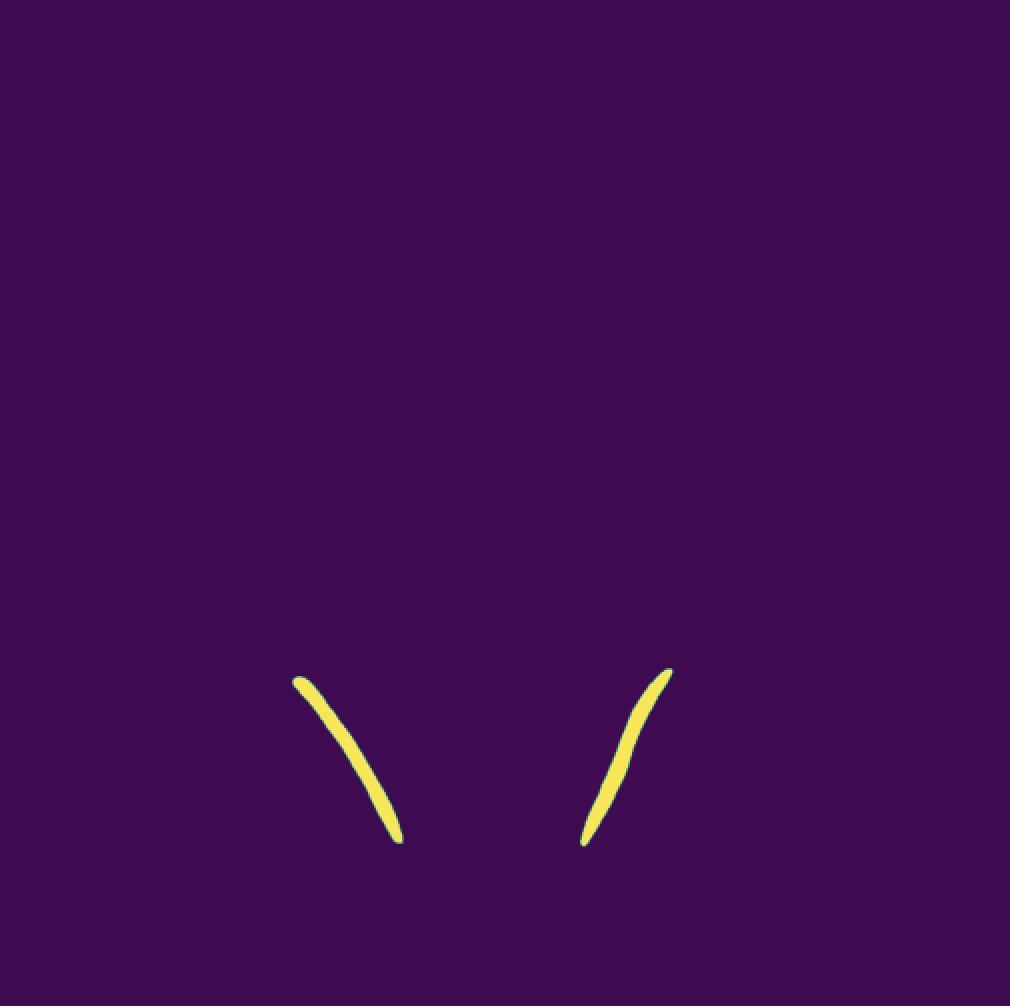}
                \caption{GT SNCD}
                \label{fig:sncd_gt}
        \end{subfigure}%
        \begin{subfigure}[b]{.2\textwidth}
                \includegraphics[width=.75\linewidth]{Images/Pred_SNCD-1.png}
                \caption{pred SNCD}
                \label{fig:sncd_pred}
        \end{subfigure}
        
        \begin{subfigure}[b]{.2\textwidth}
                \includegraphics[width=.75\linewidth]{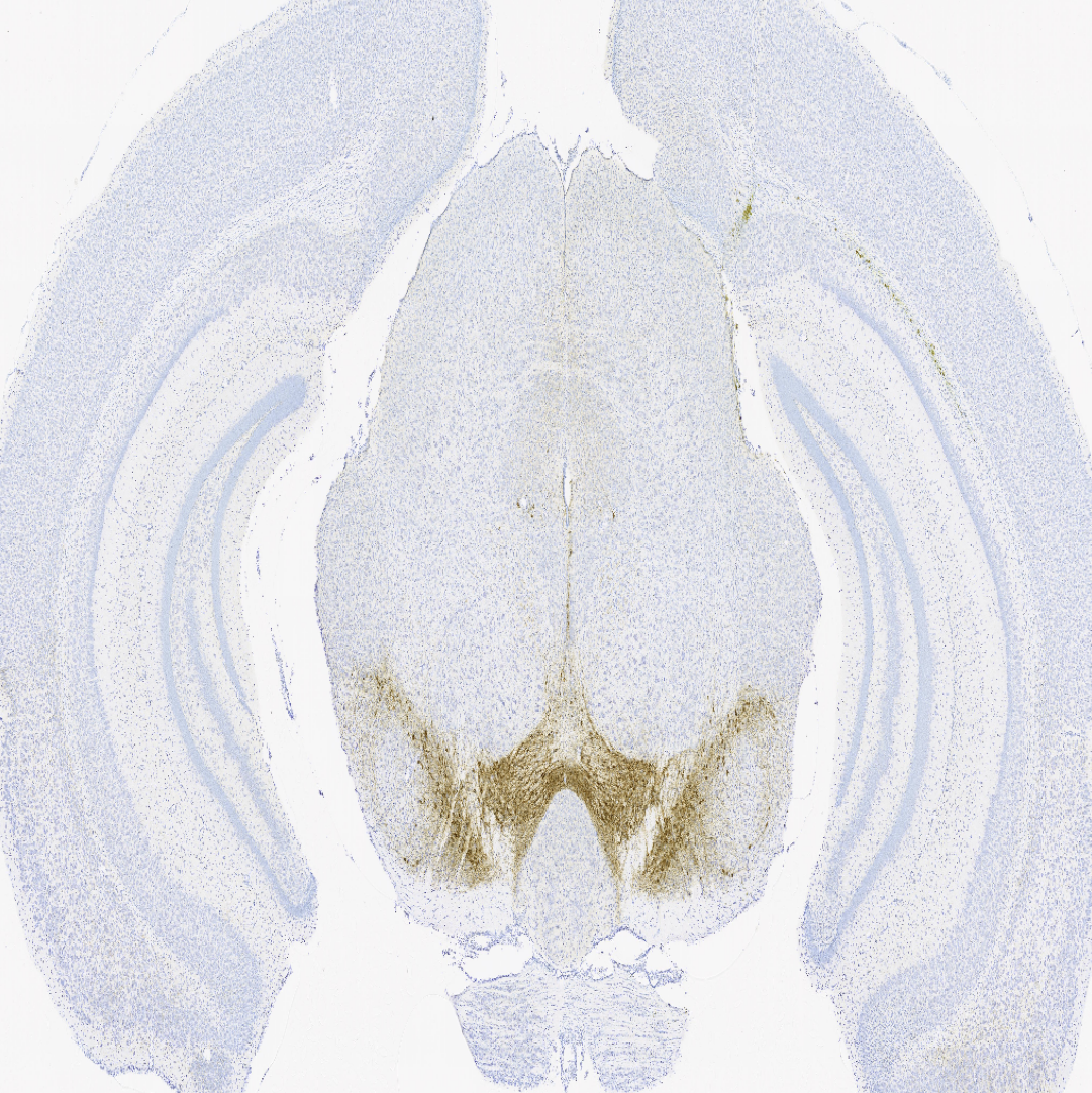}
                \caption{Image 2}
                \label{fig:image_test_2}
        \end{subfigure}%
        \begin{subfigure}[b]{.2\textwidth}
                \includegraphics[width=.75\linewidth]{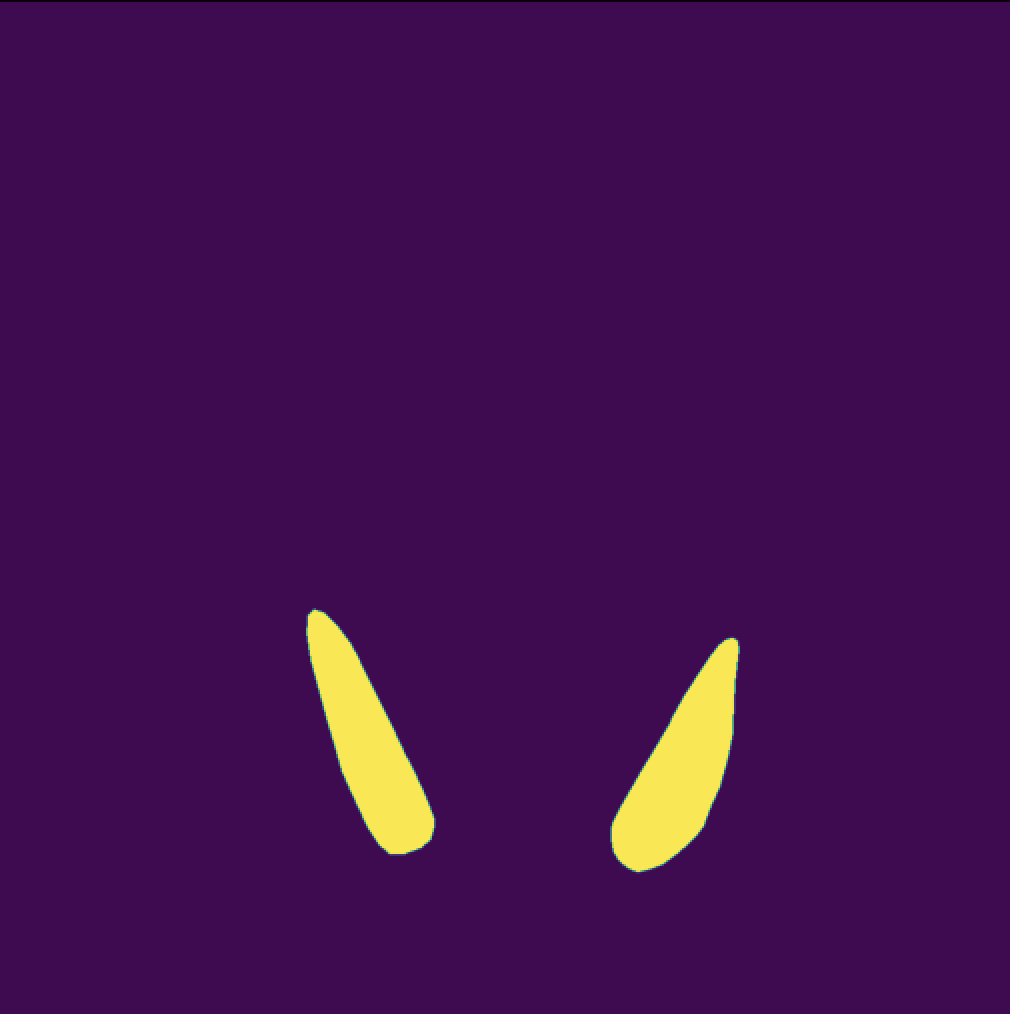}
                \caption{GT SNr}
                \label{fig: SNr_gt_2}
        \end{subfigure}%
        \begin{subfigure}[b]{.2\textwidth}
                \includegraphics[width=.75\linewidth]{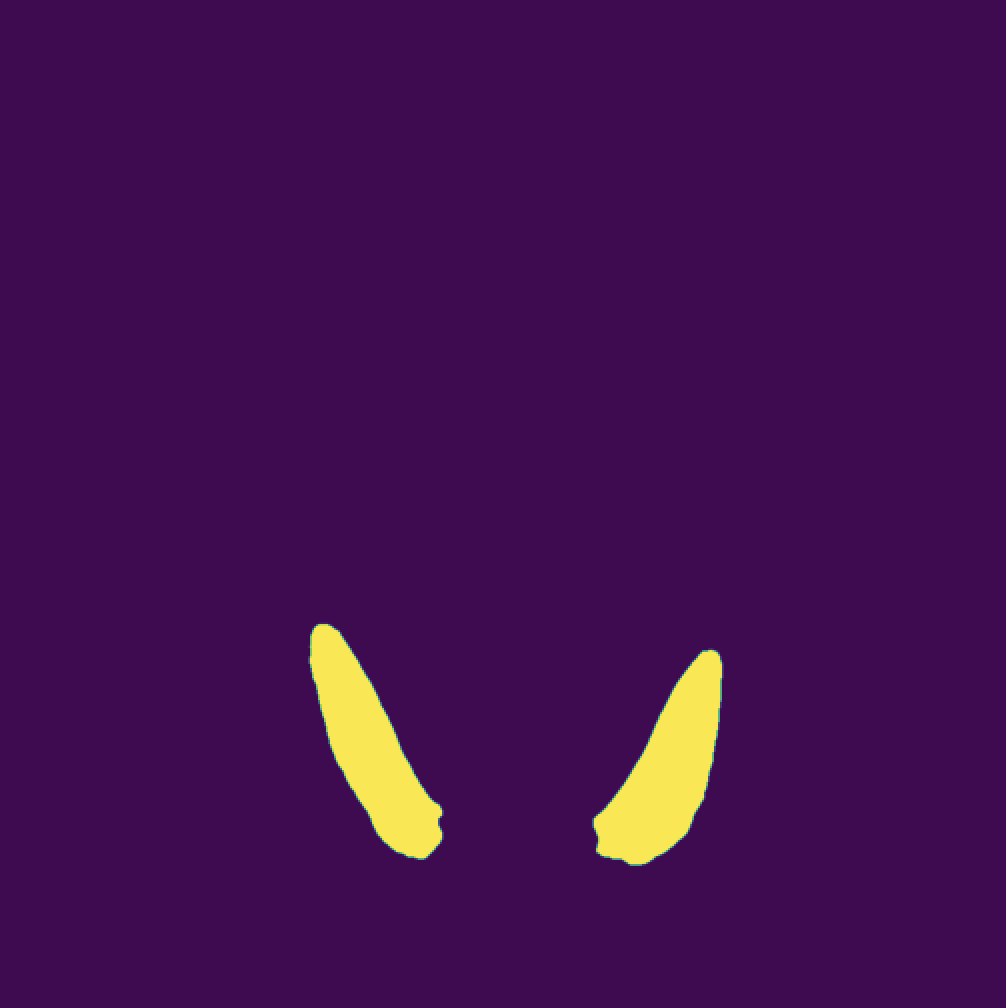}
                \caption{pred SNr}
                \label{fig:pred_SNr_2}
        \end{subfigure}%
        \begin{subfigure}[b]{.2\textwidth}
                \includegraphics[width=.75\linewidth]{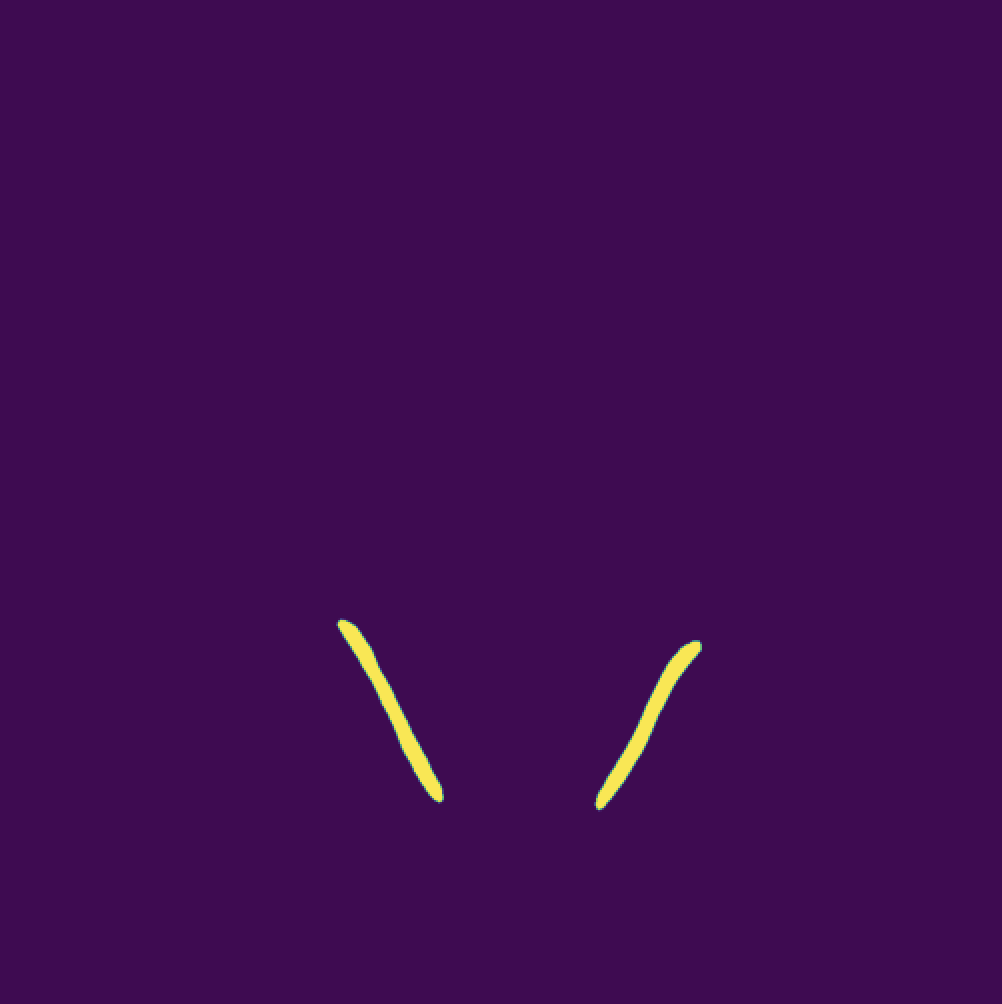}
                \caption{GT SNCD}
                \label{fig:sncd_gt_2}
        \end{subfigure}%
        \begin{subfigure}[b]{.2\textwidth}
                \includegraphics[width=.75\linewidth]{Images/Pred_SNCD-2.png}
                \caption{pred SNCD}
                \label{fig:sncd_pred_2}
        \end{subfigure}
        \caption{Two image example from the test set: Including ground truth(GT) SNr \& SNCD with their corresponding predictions(pred).}\label{fig:predictions}
\end{figure}


\begin{figure}
    \centering
    \includegraphics[width=.9\linewidth]{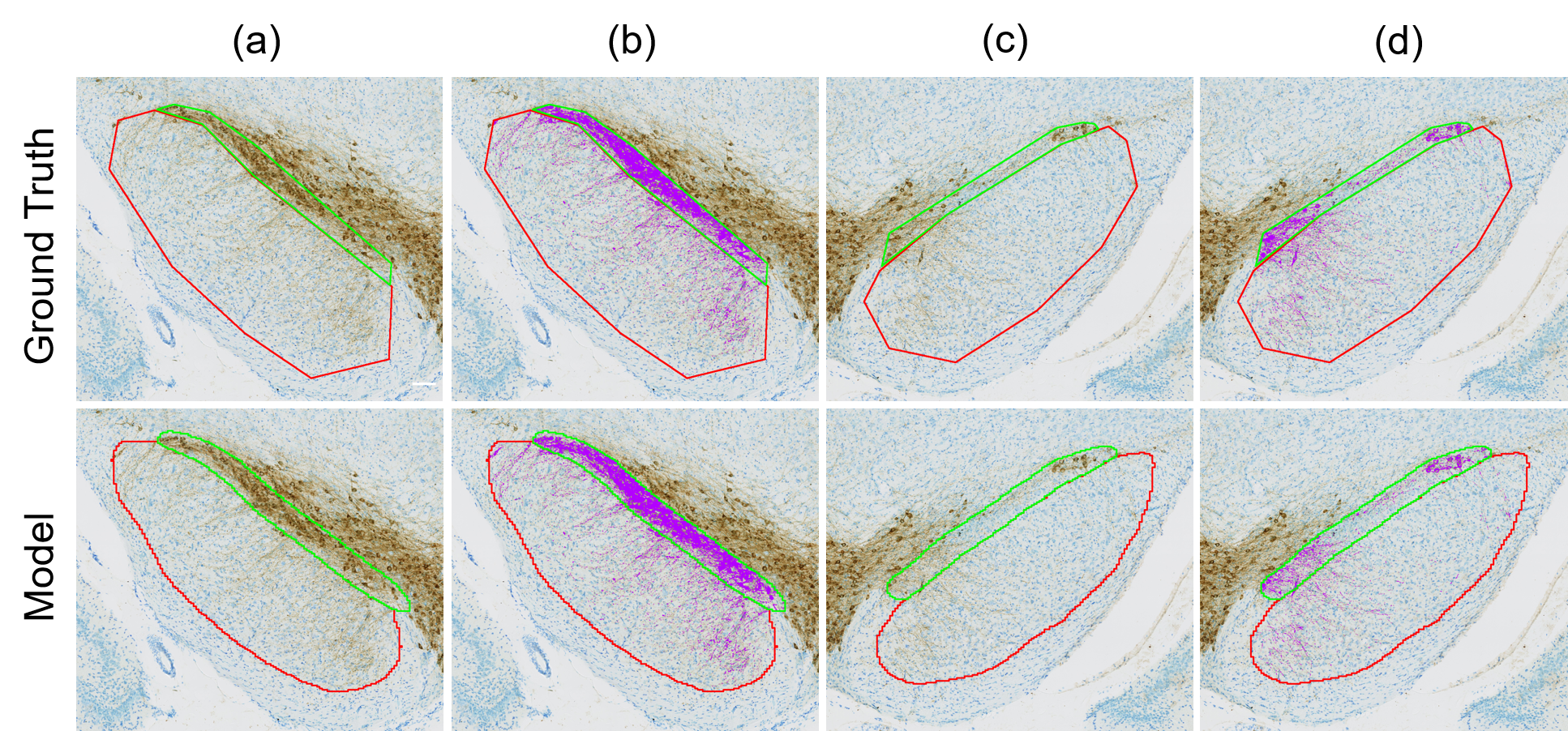}
    \caption{TH optical density measurement examples (b,d-purple color) in SNr and SNCD regions in left (a,b) and right (c,d) side of the mouse brain. The sections are representative images from blind-test dataset. The red annotation line depicts SNr and the green annotation line depicts the SNCD region of the SN segmented. The purple color indicates the detection of TH staining (brown color). Nissl staining in blue indicates tissue area. Scale bar, 100 micron.}
    \label{fig:TH_OD}
\end{figure}

\begin{figure}
    \centering
    \includegraphics[width=.9\linewidth]{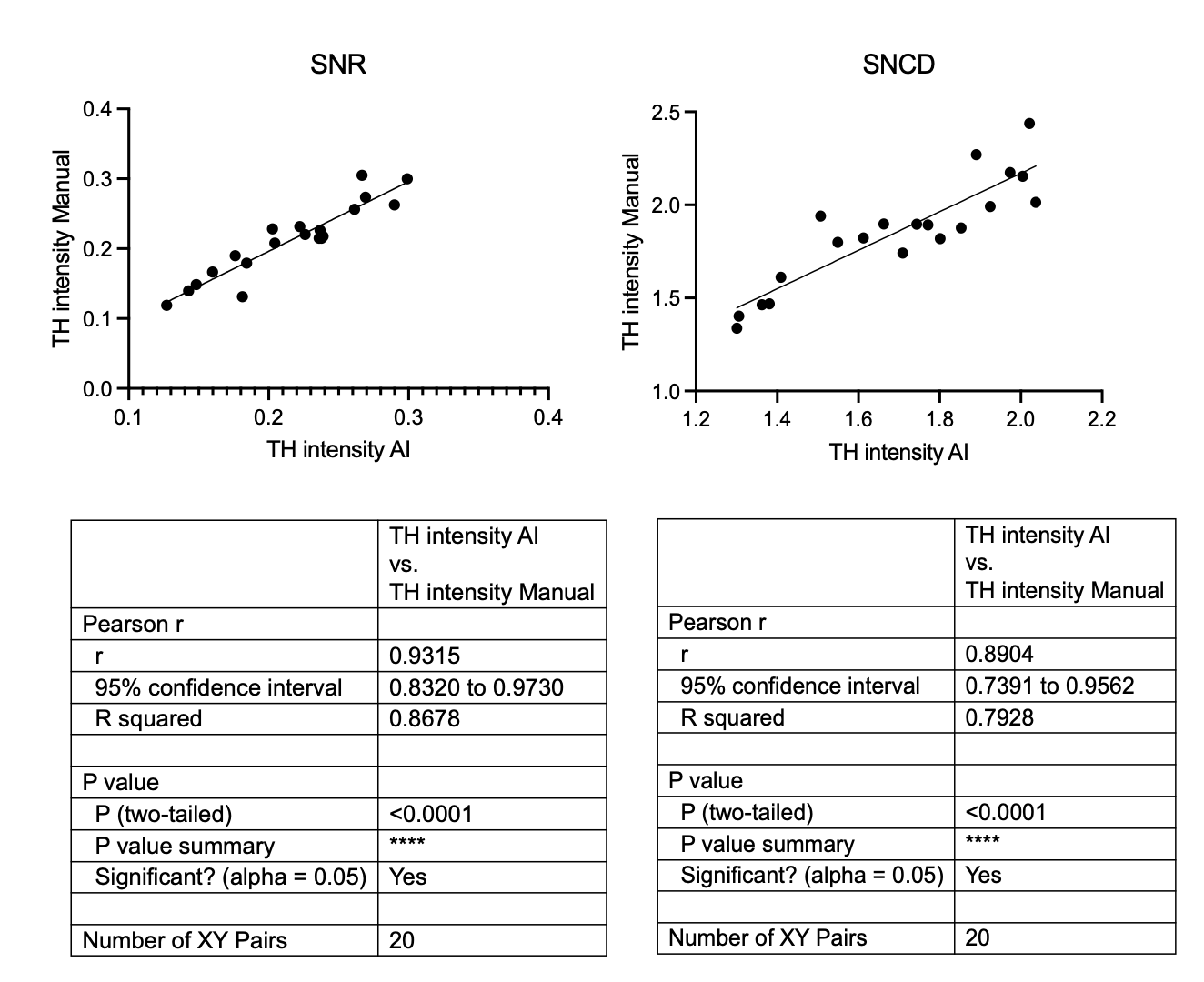}
    \caption{Correlation between TH intensity (normalized optical density) from manual segmentation and segmented area by the model. The data was calculated from random 20 mice hemi-brains stained with TH and segmented by neuroscientist (ground truth) and model (blind-test dataset).}
    \label{fig:correlation}
\end{figure}



As shown in \autoref{fig:predictions}, SN regions with different background containing nuclear staining (light blue- Nissl, Violet- Thionine) was effectively segmented to depict the SNr and SNCD region.
In the magnified image \autoref{fig:TH_OD}, the model was able to detect SNr and SNCD in the right brain hemisphere where TH signal was decreased (pathological trigger by stereotactic injection into the SN) further proving it's ability to segment SNr and SNCD without relying on TH signal. 
 The model segmented both the SNr and SNCD in both brain hemispheres. A neuropathology expert manually annotated the sections for SNr and SNCD in parallel to compare with the model segmentation. \autoref{fig:correlation} shows the TH optical density depicting DA neuronal health measured in both manually annotated and AI model segmented samples in the SNr and SNCD sub-regions. The $R^{2}$ for TH intensity between the manual segmentation and model generated segmentation for SNR is 0.8678 and SNCD is 0.7928 respectively. Considering the  diversity and complexity of biological data, a positive correlation of $\approx 0.85$ holds very high significance and confidence on the model. This data clearly shows a strong correlation between the manual segmentation and model generated segmentation. The p-value of $< 0.0001$ also shows how efficiently and precisely the model can be used to analyze the DA neuronal health in animal studies.

%% file: 05_Discussion.tex
\section{CONCLUSION}

In this paper, we were able to deploy a convolutional neural network to segment anatomical sub-region of the brain with a Dice coefficient of 87\%. The model is able to detect health status of dopamenergic neurons in SNr and SNCD which is associated with PD pathology. The model is robust enough to detect SNr and SNCD in brain sections with various background nuclear staining where the TH signal is lost. The quantitative and qualitative results together show the potential of the model and how this approach can be applied to segment other anatomical regions. To the best of our knowledge, this is one of the first machine learning based segmentation of defined anatomical sub-regions of the brain in 2D pathological sections. The fast turnover of the model also would save enormous amount of time for the biologist to make quicker decisions on drug efficacy. It also solves one of the major problem in medical imaging that arises from user (neurology expert) based associated bias. 
Applying this straightforward yet scalable method to additional anatomical regions or diseases yields remarkably promising results.  Overall, this AI model shows that machine learning-aided read out for measuring the health of dopaminergic neurons in sub anatomical regions is faster, unbiased and precise compared to manual analysis of 2D histological sections in brain.
